\documentclass[twocolumn,showpacs,preprintnumbers,amsmath,amssymb]{revtex4}
\usepackage[dvips]{graphicx}
\usepackage{dcolumn}

\def\clap#1{\hbox to 0pt{\hss#1\hss}}

\begin{document}

\title{Single-particle density matrix for a time-dependent 
strongly interacting one-dimensional Bose gas}

\author{R.~Pezer}
\email{rpezer@phy.hr}
\affiliation{Faculty of Metallurgy, University of Zagreb,
44103 Sisak, Croatia}
\author{T.~Gasenzer}                      
\affiliation{Institut f\"ur Theoretische Physik,
Universit\"at Heidelberg, Philosophenweg 16, 69120 Heidelberg, Germany}
\author{H.~Buljan}
\affiliation{Department of Physics, University of Zagreb, 
PP 332, Zagreb, Croatia}

\date{\today}

\begin{abstract}
We derive a $1/c$-expansion for the single-particle density matrix
of a strongly interacting time-dependent one-dimensional Bose gas,  
described by the Lieb-Liniger model ($c$ denotes the strength of 
the interaction). The formalism is derived by expanding Gaudin's Fermi-Bose 
mapping operator up to $1/c$-terms. 
We derive an efficient numerical algorithm for calculating the density matrix 
for time-dependent states in the strong coupling limit, which evolve from a 
family of initial conditions in the absence of an external potential. 
We have applied the formalism to study contraction dynamics of a localized 
wave packet upon which a parabolic phase is imprinted initially. 
\end{abstract}

\maketitle


\section{Introduction \label{sec:intro}}

One of the most attractive many-body quantum systems nowadays has been
introduced by Lieb and Liniger in their landmark paper more than forty years 
ago \cite{Lieb1963}. The system is composed of $N$ identical 
$\delta$-interacting bosons in one spatial dimension and is referred to 
as a Lieb-Liniger (LL) gas. They have presented an explicit form of the 
many-body wave function for a homogeneous gas with periodic boundary conditions
\cite{Lieb1963}, including equations describing the ground state and the 
excitation spectrum. 
In the strongly interacting "impenetrable-core"  regime \cite{Girardeau1960}, 
such a one dimensional (1D) system is referred to as the Tonks-Girardeau (TG) 
gas; exact solutions in this limit are obtained by Girardeau's Fermi-Bose 
mapping \cite{Girardeau1960}. 
Following Ref. \cite{Lieb1963}, Yang and Yang \cite{Yang1969} have eliminated a 
possible existence of phase transitions in the LL system by proving analyticity 
of the partition function.
After many recent experimental successes \cite{OneD,Fertig2005,TG2004,
Kinoshita2006,Hofferberth2007,Amerongen2008}
in realization of effectively one dimensional (1D) interacting gases, from the 
weak up to the strongly interacting TG regime \cite{TG2004}, the LL model has 
attracted 
considerable attention of the physics community. 
There is a clear reason for this; nontrivial quantum many-body systems are 
notoriously oblique to a quantitative analysis, and therefore possibility of 
an exact treatment in particular cases, together with experimental realization, 
is of great value. Moreover, exact solutions can be useful as a 
benchmark for approximate treatments aiming to describe a broader 
range of physical systems. 

Even though exact LL many-body wave functions can be constructed in some 
cases (e.g., stationary 
\cite{Lieb1963,McGuire1964,Gaudin1971,Muga1998,Sakmann2005,Batchelor2005,
Kinezi2006,Sykes2007} or time-dependent wave functions \cite{Gaudin1983,
Girardeau2003,Buljan2008,Jukic2008}), the calculation of observables 
(correlation functions) from such solutions usually poses a major difficulty in 
practice \cite{Korepin1993,Jimbo1981,Olshanii2003,Gangardt2003,
Astrakharchik2003,Kheruntsyan2005,Brand2005,Forrester2006,Caux2007,
Calabrese2007,Imambekov2008,Gritsev2009}. 
Various methods have been employed to overcome this difficulty including, 
for example, the quantum inverse scattering method (e.g., see 
\cite{Korepin1993,Caux2007}) and quantum Monte Carlo integration 
\cite{Astrakharchik2003}.
A recent discussion of several exact methods for the calculation of correlation
functions of a nonequilibrium 1D Bose gas can be found in 
Ref.~\cite{Gritsev2009}. In the TG limit, the momentum distribution can be 
analytically studied for a ring geometry, and also for harmonic confinement 
(e.g., see \cite{Lenard1964,Forrester2003}). Numerical methods for the 
calculation of the reduced single-particle density matrix 
(RSPDM) can be performed efficiently for various TG states (ground state, 
excited and time-dependent states, see Ref.~\cite{Rigol2005} for hard-core 
bosons on the lattice, and Ref.~\cite{Pezer2007} for the continuous 
TG model \cite{Girardeau1960}).

Ultracold atoms in 1D atomic wave guides enter the strongly interacting 
regime at low temperatures, in tight transverse confinement, and with 
strong effective interactions \cite{Olshanii,Petrov,Dunjko}. 
The correlations functions of a LL gas can in this limit be calculated by using 
$1/c$ expansions (e.g., see \cite{Jimbo1981,Gangardt2003,Brand2005}) from the 
TG $(c\rightarrow \infty)$ regime. These calculations in 
the strongly interacting limit exploit the fact that a bosonic LL gas 
is dual to a fermionic system \cite{Cheon1999}, such that weakly 
interacting fermions correspond to strongly interacting bosons and vice versa 
\cite{Cheon1999,Yukalov2005}. A strongly interacting 1D Bose gas was studied in 
Ref.~\cite{Sen2003} by using perturbation theory for the dual fermionic system. 
In Ref.~\cite{Brand2005}, the dynamic structure factor was calculated for zero 
and finite temperatures.

Here we calculate the $1/c$ correction for the RSPDM of a Lieb-Liniger 
gas, which adds upon a recently obtained formula for the RSPDM of 
a TG gas \cite{Pezer2007}. The method is derived by using the $1/c$ term
of the so-called Fermi-Bose (FB) mapping operator introduced by Gaudin 
\cite{Gaudin1983}. 
The FB operator method provides us with exact time-dependent solutions 
of a LL model \cite{Gaudin1983,Buljan2008} in the absence of external 
potentials and other boundary conditions; it was recently used to study 
free expansion of a LL gas \cite{Jukic2008}. 
We derive an efficient numerical algorithm for calculating the RSPDM 
for time-dependent states in the strong coupling limit, which evolve from a 
family of initial conditions in the absence of an external potential. 
We employ it to study the evolution of a many-body wave packet with a 
parabolic phase imprinted at $t=0$, which corresponds to focusing with a lens
in optics, a technique experimentally feasible in 1D atomic gases 
\cite{Amerongen2008}. 

This work complements the studies of nonequilibrium dynamics of interacting 
Bose gases which have been addressed by use of the LL model away from
\cite{Girardeau2003,Buljan2008,Jukic2008,Gritsev2009} and in the TG limit 
\cite{Girardeau2000,Girardeau2000a,Rigol2005,Minguzzi2005,DelCampo2006,
Rigol2006,Buljan2006,Pezer2007,Gangardt2007}. The phenomenological 
relevance of these studies is underlined by the fact that nonequilibrium 
dynamics is accessible experimentally \cite{Kinoshita2006,Hofferberth2007}.

The paper is organized as follows: Section \ref{sec:math} gives a 
detailed account of the formalism, where we outline the procedure 
for the calculation of the RSPDM. In section \ref{sec:neqHO} we present the 
example of a Bose gas initially in the ground state within a harmonic trap, 
effectively in the TG regime; at $t=0$ a parabolic phase is applied on the 
initial state and the harmonic potential is turned off. The subsequent dynamics
leads to focusing of the cloud and local increase of the $1/c$ correction. 
Details of the derivation and relevant mathematical 
identities are collected in the Appendices.

\section{Formalism for the 
time-dependent strongly-interacting Lieb-Liniger gas
\label{sec:math} }

The Lieb-Liniger model describes a system of bosons interacting 
via pointlike interactions; the many-body 
Schr\"odinger equation for the LL gas of $N$ such bosons reads \cite{Lieb1963}
\begin{equation}
i \frac{\partial \psi_{LL}}{\partial t}=
-\sum_{i=1}^{N}\frac{\partial^2 \psi_{LL}}{\partial x_i^2}+
\sum_{1\leq i < j \leq N} 2c\,\delta(x_i-x_j)\psi_{LL}.
\label{eq:LLmodel}
\end{equation}
Here, $\psi_{LL}(x_1,\ldots,x_N,t)$ is the time-dependent bosonic wave function,
$c$ is the strength of the interaction. 
For now, we assume the absence of any external potential 
and boundary conditions. Under these circumstances, 
the time-dependent LL model \eqref{eq:LLmodel} can be solved by employing 
the Fermi-Bose transformation \cite{Gaudin1983,Buljan2008}. 
This method can be applied, e.g., to exactly study free expansion
from an initially localized state \cite{Jukic2008}. 
If $\psi_F(x_1,\ldots,x_N,t)$ is an antisymmetric (fermionic)
wave function, which obeys the Schr\" odinger equation
for a noninteracting Fermi gas,
\begin{equation}
i \frac{\partial \psi_F}{\partial t}=
-\sum_{i=1}^{N}\frac{\partial^2 \psi_F}{\partial x_i^2},
\label{SchF}
\end{equation}
then the wave function
\begin{multline}
\psi_{LL}(x_1,\ldots,x_N,t) = \sqrt{{\mathcal N}(c)} 
\!\!\prod_{1\leq r < l \leq N} \!\!
\Big[
\mbox{sgn}(x_l-x_r)+ \\ \frac{1}{c}
\left(
\frac{\partial}{\partial x_{l}}-
\frac{\partial}{\partial x_{r}}
\right)
\Big] \psi_F(x_1,\ldots,x_N,t)
\label{eq:ansatz}
\end{multline}
obeys Eq.~\eqref{eq:LLmodel} as pointed out in Ref.~\cite{Gaudin1983}
(see also Refs.~\cite{Buljan2008,Jukic2008}). 
Here, \({\mathcal N}(c)\) is a normalization constant, and the 
differential operator in the square brackets denotes the Fermi-Bose 
mapping operator \cite{Buljan2008}. 
The derivatives do not act on any of the sign functions (the sign functions 
can be avoided by working in only one sector of the configuration space, e.g., 
for $x_1<x_2\ldots<x_N$, e.g., see Refs. \cite{Buljan2008,Jukic2008}).
We assume that \(\psi_F(x_1,\ldots,x_N,t)\) is normalized to unity. 
Equation~\eqref{eq:ansatz} can be reorganized in a finite power
series with terms of order \(1/c^m\), where $m=0,1,\ldots,N(N-1)/2$.  
By using the Fermi-Bose mapping operator in this form, 
one obtains a systematic expansion of the exact many-body 
wave function $\psi_{LL}$ in the inverse interaction strength $1/c$.  

The expansion of the wave function~\eqref{eq:ansatz} in the inverse coupling 
strength $1/c$ is particularly useful in the strong coupling limit,
as it allows an approximate calculation of one-body observables 
contained within the reduced single-particle density matrix (RSPDM). 
In the Tonks-Girardeau limit, where $c=\infty$, a formula for efficient 
calculation of the RSPDM has recently been derived \cite{Pezer2007}. 
Here we generalize that result to include the $1/c$ term in the expansion. 
By keeping only the $1/c$ terms in the expansion, Eq.~\eqref{eq:ansatz}
reduces to 
\begin{multline}
\psi_{LL}(x_1,\ldots,x_N,t) \simeq
\sqrt{{\mathcal N}(c)} \! 
\!\!\prod_{1\leq i < j \leq N} \! \mbox{sgn}(x_j-x_i) 
\Big[ \psi_F + \\ \frac{1}{c} \sum_{1\leq r < l \leq N} 
\mbox{sgn}(x_l-x_r)\left( \frac{\partial \psi_F}{\partial x_l} 
- \frac{\partial \psi_F}{\partial x_r} \right)
\Big]_{(x_1,\ldots,x_N,t)}
\label{eq:wfEx}
\end{multline}
The first term is simply the TG gas wave function \cite{Girardeau1960},
while the second term gives the \(1/c\) correction to the TG wave function 
when the coupling constant \(c\) is finite.
This expression is the starting point for all results that will be derived
in this paper. The RSPDM is defined as 
\begin{align}
\rho_{LL} (x,y,t) = 
&N \!\int \!\! dx_2 \cdots dx_N\, \psi_{LL}(x,x_2,\ldots,x_N,t)^{*} 
\nonumber \\
&\times \psi_{LL}(y,x_2,\ldots,x_N,t).
\label{eq:rhoLL}
\end{align}
On inserting Eq.~\eqref{eq:wfEx} into Eq.~\eqref{eq:rhoLL}) we obtain 
a formal expression for the \({\cal O}(1/c)\) correction of the RSPDM:
\begin{equation}
\label{eq:rspdmExA}
\begin{aligned}
 \rho_{LL}(x,y,t) 
 =&\,\, {\mathcal N}(c) \hat{I}(X) \, \psi_F^* (x,X,t) \psi_F (y,X,t) \\ 
 & +\ \frac{1}{c} {\mathcal N}(c)\! \sum_{1\leq r < l \leq N} \!\hat{I}(X)
 \Big[ \mbox{sgn}(x_l-x_r) \\
 & \left( \frac{\partial \psi_F}{\partial x_l} 
 - \frac{\partial \psi_F}{\partial x_r} \right)_{(x,X,t)}^* \psi_F (y,X,t) \\ 
 & +\ \mbox{sgn}(x_l-x_r)\,\psi_F^* (x,X,t) \, \\
 & \left( \frac{\partial \psi_F}{\partial x_l} 
 - \frac{\partial \psi_F}{\partial x_r} \right)_{(y,X,t)}
 \Big]   + \mathcal{O}(1/c^2). 
\end{aligned}
\end{equation}
Here, \(X=(x_2,\ldots,x_N)\) and the integral operator $\hat{I}(X)$ is 
defined as:
\begin{equation} \label{eq:intop}
\hat{I}(X) = N \prod_{n=2}^{N} \int_{-\infty}^{\infty} \!\! dx_n \, 
\mbox{sgn}(x-x_n)\,\mbox{sgn}(y-x_n).
\end{equation}
The first term on the right hand side of Eq. \eqref{eq:rspdmExA} is the TG gas
RSPDM \cite{Pezer2007}. 
It can be proven (as is done in Appendix \ref{app:formulas}) that 
only the
partial derivatives with respect to the first coordinate \(x_1\) in 
Eq.~\eqref{eq:rspdmExA} 
give a nonvanishing contribution. After eliminating the vanishing terms 
from Eq.~\eqref{eq:rspdmExA} we are left with 
\begin{align}
\rho_{LL}(x,y,t) & = {\mathcal N}(c) \rho_{TG}(x,y,t) + 
\frac{1}{c} {\mathcal N}(c) 
\nonumber \\ 
\times \sum_{l=2}^N & \hat{I}(X) 
\Big[ 
\mbox{sgn}(x-x_l)
\left( \frac{\partial \psi_F}{\partial x_1} \right)_{(x,X,t)}^* \!\!\! 
\psi_F (y,X,t) + 
\nonumber \\
\mbox{sgn}&(y- x_l) \, \psi_F^* (x,X,t) 
\left( \frac{\partial \psi_F}{\partial x_1}  \right)_{(y,X,t)} 
\Big] + \mathcal{O}(1/c^2) \nonumber \\
= \rho_{TG}&(x,y,t) + \frac{1}{c}  [\eta(x,y,t)+\eta(y,x,t)^*]
+ \mathcal{O}(1/c^2).
\label{eq:rspdmEx}
\end{align}
In Eq. \eqref{eq:rspdmEx} we have implicitly defined the quantity 
$\eta(x,y,t)$:
\begin{flalign} 
\eta(x,y,t) = \sum_{l=2}^{N} \hat{I}(X) \, \mbox{sgn}(x-x_l)
\left( \frac{\partial \psi_F}{\partial x_1} \right)_{(x,X,t)}^* 
\!\!\! \psi_F (y,X,t) \nonumber \\
 = (N-1) \hat{I}(X) \, \mbox{sgn}(x-x_2)
\left( \frac{\partial \psi_F}{\partial x_1} \right)_{(x,X,t)}^* 
\!\!\! \psi_F (y,X,t).
\label{eq:auxterm}
\end{flalign}
By using the calculation presented in Appendix \ref{app:formulas} 
it follows that  
\[
\int dx [\eta(x,x,t)+\eta(x,x,t)^*]=0,
\]
that is, the $1/c$ correction to the single-particle density $\rho_{LL}(x,x,t)$ 
increases the density in some regions of space, but also lowers it in others
such that the integral over the terms of order $1/c$ is zero. By using this
result we see that in the leading $1/c$ approximation we can take 
\({\mathcal N}(c)=1\); this fact has already been utilized in the 
last line of Eq. \eqref{eq:rspdmEx}.
It is straightforward to verify that the integrals 
$\hat{I}(X) \mbox{sgn}(x-x_l)
\left( \frac{\partial \psi_F}{\partial x_1} \right)_{(x,X)}^* \psi_F (y,X)$ 
are independent of $l$ ($2\leq l \leq N$), which yields the second identity 
in Eq. \eqref{eq:auxterm}. 
The last line of Eq.~\eqref{eq:rspdmEx} verifies that 
to order $1/c$ the RSPDM possesses as required the symmetry 
$\rho_{LL}(x,y,t)=\rho_{LL}(y,x,t)^*$.  

Up to this point we did not make any assumptions on the structure of 
$\psi_F$ (except that it is antisymmetric and normalized), that is,
the derivation was general, valid even if the wave functions 
should describe the gas in an external potential etc. 
Quite generally, $\psi_F$ can also be considered as a function of $c$,
and one could expand it in powers of $1/c$.
>From this point on, $\psi_F$ will be represented as a Slater determinant formed from 
single particle wave functions,
\begin{equation}
\psi_F=\frac{1}{\sqrt{N!}}\sum_{P}(-)^P
\phi_{P1}(x_1,t)\cdots \phi_{PN}(x_N,t);
\label{eq:psiFdet}
\end{equation}
such wave functions can be used to study dynamics on an infinite line,
which arises from initial conditions given by Eqs. \eqref{eq:wfEx}
and \eqref{eq:psiFdet}. 
In Eq. \eqref{eq:psiFdet}, $\phi_{j}(x,t)$ ($j=1,\ldots,N$) are orthonormal 
single-particle wave functions which obey the Schr\"odinger equation 
$i \partial \phi_{j}/ \partial t=- \partial^2 \phi_{j}/ \partial x^2$.
$P$ denotes a permutation $P(1,\ldots,N)=(P1,\ldots,PN)$ of the particle number 
indices, and $(-)^P$ is its signature. 
In this form $\psi_F$ enables us to rewrite Eq.~\eqref{eq:rspdmEx} 
such that it involves 1D integrals only, resulting in certain 
algebraic cofactors suitable for numerical calculation.

\subsection{Algorithm for RSPDM calculation}

For the sake of clarity of the presentation, we first describe the 
algorithm for calculation of the RSPDM, and only afterwards 
provide its derivation. Without loss of generality we consider the 
case \(x<y\). The first step is to calculate the integrals 
\begin{equation}
I_{k,l}(x,y,t) = \delta_{kl}-2\!\!\int_{x}^{y} 
\!\!\! dx' \, \phi_{k}^{*}(x',t)\phi_{l}(x',t)
\label{eq:fint}
\end{equation}
and                       
\begin{equation}
\overline{I}_{k,l}(y,t) = -\delta_{kl}+2\!\!\int_{-\infty}^{y} 
\!\!\!\!\! dx' \, \phi_{k}^{*}(x',t)\phi_{l}(x',t);
\label{eq:fintD}
\end{equation}
for $k,l=1,\ldots,N$.
\begin{widetext}
These integrals are arranged in the following matrices:
\begin{equation}
{\mathbf P}(x,y,t) = 
\left[ \begin{array}{cccc}
    I_{1,1}(x,y,t) & I_{1,2}(x,y,t) 
                    & \ldots & I_{1,N}(x,y,t) \\
    I_{2,1}(x,y,t) & I_{2,2}(x,y,t) 
                    & \ldots & I_{2,N}(x,y,t) \\
    \vdots & \vdots & \ddots & \vdots \\
    I_{N,1}(x,y,t) & I_{N,2}(x,y,t) 
                    & \ldots & I_{N,N}(x,y,t)
    \end{array}\right],
\label{eq:AmatDP}
\end{equation}
and 
\begin{equation}
{\mathbf P}^{(l)}(x,y,t) = 
\left[ \begin{array}{cccccc}
    I_{1,1}(x,y,t) & I_{1,2}(x,y,t) 
                    & \ldots & \overline{I}_{1,l}(y,t) 
		    & \ldots & I_{1,N}(x,y,t) \\
    I_{2,1}(x,y,t) & I_{2,2}(x,y,t) 
                    & \ldots & \overline{I}_{2,l}(y,t) 
		    & \ldots & I_{2,N}(x,y,t) \\
    \vdots & \vdots &  & \ddots & & \vdots \\
    I_{N,1}(x,y,t) & I_{N,2}(x,y,t) 
                    & \ldots & \overline{I}_{N,l}(y,t) 
		    & \ldots & I_{N,N}(x,y,t)
    \end{array}\right].
\label{eq:AmatD}
\end{equation}
\end{widetext}

Let us define the column vector 
\begin{equation}
{\mathbf \Psi }(x,t)=
\begin{pmatrix}
\phi_{1}(x,t) \\ \vdots \\ \phi_{N}(x,t)
\end{pmatrix},
\end{equation}
and its first spatial derivative
\begin{equation}
{\mathbf \Psi }'(x,t)=
\begin{pmatrix}
\phi_{1}'(x,t) \\ \vdots \\ \phi_{N}'(x,t)
\end{pmatrix}.
\end{equation}
The TG reduced single-particle density matrix is given by \cite{Pezer2007}
\begin{multline}
\rho_{TG}(x,y,t)= 
\det\left[ {\mathbf P}(x,y,t) \right] \\
{\mathbf \Psi }^{\dag}(x,t)
\left[ {\mathbf P}(x,y,t)^{-1} \right]^{T}
{\mathbf \Psi }(y,t).
\end{multline}
\begin{widetext}
The quantity $\eta(x,y,t)$ can also be written in a convenient matrix form 
(suitable for efficient numerical implementation):
\begin{align}
\eta(x,y,t) =
 \sum_{l=1}^{N} \left\{ \det\left[ {\mathbf P}^{(l)}(x,y,t) \right]
 {\mathbf \Psi }'^{\dag}(x,t)
 \left[{\mathbf P}^{(l)}(x,y,t)^{-1} \right]^{T}
 {\mathbf \Psi }(y,t) \right\} \nonumber \\
 \quad-\ \det\left[ {\mathbf P}(x,y,t) \right] 
 {\mathbf \Psi }'^{\dag}(x,t)
 \left[{\mathbf P}(x,y,t)^{-1} \right]^{T}
 {\mathbf \Psi }(y,t) \label{eq:eta-num}
\end{align}
\end{widetext}
If any of the matrices \({\mathbf P}^{(l)}\) happen to be singular, we can 
resort to a direct calculation via algebraic cofactors (see the proof of 
the algorithm in Appendix \ref{app:determinant}). This happens rarely and 
only for some particular high-symmetry points. Equations 
\eqref{eq:AmatDP}--\eqref{eq:eta-num} provide the grounds for an efficient 
numerical method for calculating the RSPDM, which is a generalization of the 
previously introduced method for the TG gas \cite{Pezer2007}.                

It is convenient to calculate the diagonal correction to the RSPDM. 
In this case, the matrices appearing in Eq.~\eqref{eq:eta-num} are
very simple and it is straightforward to obtain
\begin{multline}
c \left[\frac{\rho_{LL}(x,x,t)}{{\mathcal N}(c)} -  
\rho_{TG}(x,x,t)\right] \simeq \\
\left[
Tr (\rho') \, Tr (\overline{I}) -
Tr(\rho' \cdot \overline{I} ) 
\right]_{(x,t)}
\label{eq:eta-diag}
\end{multline}
where on the right hand side \(\rho\) is \(N\times N\)
matrix given by
\[
\rho_{k,l}(x,t)= \phi_{k}^{*}(x,t)\phi_{l}(x,t)
\]
and primes denote the first spatial derivative. We can also omit the 
normalisation constant from Eq.~\eqref{eq:eta-diag} since 
\({\mathcal N}(c)=1\) within the $1/c$ approximation considered here.

\section{Contraction dynamics of strongly interacting 1D Bose gas 
         \label{sec:neqHO}}

In this section we apply the presented formalism to investigate 
time-evolution of a many-body wave packet which contracts in space after a 
parabolic phase was imprinted on it. 
More specifically, we consider $N$ strongly interacting 
bosons which are initially in the ground state in the presence of a harmonic 
potential. We assume that this initial state is well described by the 
Tonks-Girardeau ground state wave function, a symmetrized Slater determinant. 
As described in more detail below, at time $t=0$, a parabolic phase is 
suddenly imprinted onto this state, and the harmonic potential is turned off.
Imprinting the parabolic phase is experimentally feasible \cite{Amerongen2008} 
and is equivalent to the action of a focusing lens on the travelling-wave 
state of a light beam in optics. 
In the context of atomic gases this can be achieved by applying, over a short 
period of time, a tight longitudinal harmonic potential to the 
one-dimensional gas. This results in providing different initial momenta to 
different parts of the wave packet, with the momenta being proportional to 
the distance from the center of the wave packet and directed towards the
center.

In order to connect the model written in Eq. \eqref{eq:LLmodel} to 
physical units we first write the 1D Schr\"odinger equation for this system: 
\begin{multline}
i \hbar \frac{\partial \psi_{LL}}{\partial \tau}=
\sum_{i=1}^{N} \left[ -\frac{\hbar^2}{2m}
\frac{\partial^2 }{\partial \xi_i^2}+
\frac{m \omega_{HO}^2}{2} \xi_i^2 \right]\psi_{LL} \\+
\sum_{1\leq i < j \leq N} 2g\,\delta(\xi_i-\xi_j)\,\psi_{LL},
\nonumber
\end{multline}
where $g$ is the 1D scattering length \cite{Olshanii}, $\xi$ is the 
spatial coordinate, and $\tau$ denotes time. 
Here we denote the axial trapping frequency by \(\omega_{HO}\). 
By multiplying the above expression with \( 2/\hbar\omega_{HO} \) 
we obtain a dimensionless equation suitable for numerical calculation
and a physical interpretation of the relevant scales:
\begin{multline}
i \frac{\partial \psi_{LL}}{\partial t}=
\sum_{i=1}^{N} \left[ -\frac{\partial^2 }{\partial x_i^2}+
x_i^2 \right]\psi_{LL} \\ 
+ \sum_{1\leq i < j \leq N} 2\frac{2g}{a_{HO} E_{HO}}
\,\delta(x_i-x_j)\,\psi_{LL}.
\label{eq:LLmodelTrap}
\end{multline}
where we have denoted zero point energy of the harmonic trap with
\(E_{HO}=\frac{\hbar \omega_{HO}}{2}\). The unit of time is \(2/\omega_{HO}\), 
while space is in units \(a_{HO}=\sqrt{\hbar/m\omega_{HO}}\); in other words 
\(x_i=\xi_i/a_{HO}\), and \(t= \omega_{HO} \tau/2\). If we set
\(c=2g/a_{HO} E_{HO}\) we easily verify that it is exactly 
the interaction strength of the LL model $c$, see Eq. \eqref{eq:LLmodel}.
For concreteness, let us assume that the harmonic trap has frequency 
$\omega_{HO}=60$ Hz, which is loaded with \(^{87}\)Rb atoms, $N=12$.
These values yield \(a_{HO} = 1.38 \mu\)m.
\begin{figure*}[!ht]
\centerline{
\mbox{\includegraphics[width=0.9\textwidth]{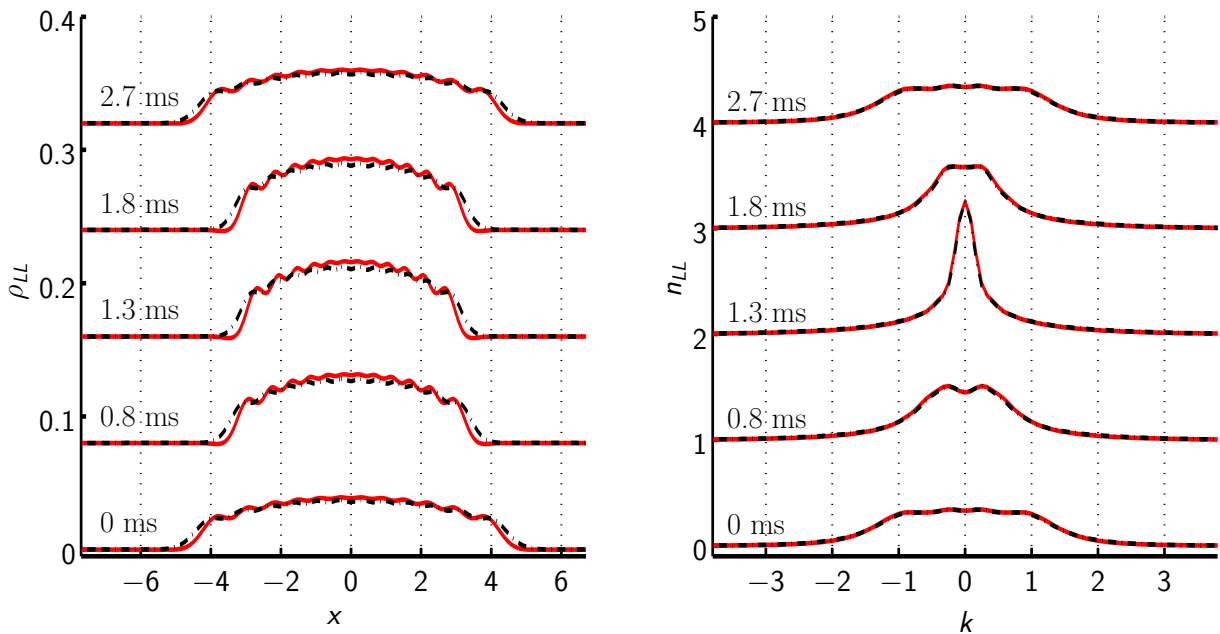}}
}                                                                      
\caption{ \label{fig01}
(color online) The $x$-space density (left, red solid line), and momentum 
distribution (right, red solid line) of the LL gas with N=12 bosons and 
\(c=50\) at different times of the propagation. At the time of 1.3 ms the
cloud attains the maximal compression.
The corresponding shapes for the TG gas are also shown as black dot-dashed 
lines. The difference between red solid and black dot-dashed lines corresponds 
to the $1/c$ correction. The curves are shifted on the ordinate axis for 
better visibility. See text for details. 
}
\end{figure*}
As we have already stated, we assume that the effective interaction strength 
$c$ is sufficiently large such that the system is initially in the 
Tonks-Girardeau regime, and that the $1/c$ correction is negligible at $t=0$. 
Under these conditions the ground state is given by 
\begin{equation}
\psi_{g.s.}=\left[\prod_{1\leq i < j \leq N} \mbox{sgn}(x_j-x_i)\right]
\det_{j,m=1}^N[\phi_j(x_m)]/\sqrt{N!},
\label{eq:inigs}
\end{equation}
where $\phi_j(x)$ is the $j$th eigenstate of the SP 
harmonic oscillator $-\phi_j''+x^2\phi_j=E_j\phi_j$. 
At $t=0$ we imprint a parabolic phase to the system. Technically, 
each of the SP wave functions in the determinant \eqref{eq:inigs} is 
multiplied by 
\[
P(x)=e^{-i (x/b)^2},
\]
where we have chosen $b=\sqrt{2}$.
Thus, the initial wave function which starts to evolve in the absence
of a trapping potential is 
\begin{multline}
\psi_{LL}(x_1,\ldots,x_N,0)\approx
\left[\prod_{1\leq i < j \leq N} \mbox{sgn}(x_j-x_i)\right] \\
\det_{j,m=1}^N[\phi_j(x_m)P(x_m)]/\sqrt{N!}.
\label{iniLL}
\end{multline}
After this initial excitation the wave packet starts to contract.
Figure \ref{fig01} displays the evolution of the SP $x$-space density 
and the momentum distribution (SP $k$-space density) for $c=50$;
the momentum distribution is defined as \cite{Lenard1964}
\[
n_{LL}(k,t) =(2\pi)^{-1} \!\int_{-\infty}^{\infty}\!\!\!\!dx 
                       \int_{-\infty}^{\infty}\!\!\!\!dx'\,
                        e^{-ik(x-x')} \rho_{LL}(x,x',t).
\]
It should be emphasized that the results are scalable, that 
is, the same results can be obtained for smaller values of $c$ 
provided that the initial state is broader (this corresponds to the 
fact that for the LL gas in equilibrium the regime is determined by
the interaction strength $c$ divided by the linear density \cite{Lieb1963}). 
As the gas becomes more dense during the contraction, the $1/c$ correction to 
the TG density becomes larger and clearly visible in the $x$-space density.
Note that the gas compresses more strongly for the finite $c$ in comparison 
to the TG results. 
After the wave packet reaches the maximally dense configuration, it starts to 
freely expand and the $1/c$ correction starts to become less visible.
It is interesting to note that while the changes in the $x$-space density are
nonnegligible in contraction, the $1/c$ correction in the $k$-space density
is practically negligible at all times. 
>From our simulations we find quite generally that the $1/c$ correction to the 
$k$-space density is much less sensitive to changes in the coupling strength
than the $x$-space density. Note that the total $x$-space density in the 
presented form of the $1/c$ approximation can become negative in some parts 
of the space, if the approximation is used beyond its range of validity 
(e.g., if c is not sufficiently large). This gives a clear indication that 
higher order terms, of order $\mathcal{O}(1/c^2)$, become relevant, at least 
in those parts of the space where the density is negative.

We emphasize again that the results presented here, for an evolution in 
one dimension without further boundary conditions, are correct up to \(1/c\) 
for all times of the evolution, i.e., there is no accumulation of the error 
in time due to the approximation and we can extract the asymptotic behaviour 
of the LL gas.

\section{Conclusion 
         \label{sec:conclusion}}

We have studied the \(1/c\) expansion in the strong coupling limit 
of the Lieb-Liniger model, exploiting the formalism based on
a mapping between free Fermi and interacting Bose gas. More specifically
we have developed an efficient algorithm to calculate the reduced one body
density matrix of the strongly interacting LL gas, which is based on the method 
previously developed for the Tonks-Girardeau gas \cite{Pezer2007}, 
and the Fermi-Bose transformation \cite{Gaudin1983,Buljan2008,Jukic2008}.
The method is suitable to describe dynamics on an infinite line, 
in the absence of external potentials, and for initial states 
which can be described by Equations \eqref{eq:wfEx} 
and \eqref{eq:psiFdet}. An attractive feature is stable accuracy for all times 
since the error is not being accumulated during the evolution. For the 
$x$-space density we provide a very compact formula Eq. \eqref{eq:eta-diag} 
that can be employed for quite large numbers of particles enabling fast 
calculation of the nonequilibrium $x$-space density dynamics.
In addition, the efficiency of the method is demonstrated 
in an example discussed in Sec. \ref{sec:neqHO}; we have examined 
nonequilibrium dynamics where a parabolic phase is imprinted onto 
a localized wave packet after which the wave packet contracts for a while. 
Our simulation studies the setup which was experimentally realized 
in Ref. \cite{Amerongen2008} for a weakly interacting Bose gas at 
finite temperature. 


\acknowledgments
R.P. and H.B. acknowledge support from the Croatian Ministry of Science
(Grant No. 119-0000000-1015). T.G. acknowledges support
by the Deutsche Forschungsgemeinschaft. This work is also
supported in part by the Croatian-German scientific collaboration
funded by DAAD and MZO\v S, and the Croatian National
Foundation for Science.

\appendix

\section{\label{app:formulas} Formulas}
             
Integrals in Eq. \eqref{eq:rspdmExA} involving derivatives in integration 
coordinates all vanish due to following expression for \(k \ne 1\)
and for any \(l \ne k\): 
\begin{align}                                                                           
& \hat{I}(X) \mbox{sgn}(x_k - x_l) 
\left( \frac{\partial \psi_F}{\partial x_k} \right)^*_{(x,X)} \psi_F (y,X)
\nonumber \\
& + \hat{I}(X) \mbox{sgn}(x_k - x_l) 
\psi_F^* (x,X) \left(\frac{\partial \psi_F}{\partial x_k} \right)_{(y,X)} 
\nonumber \\
& = \hat{I}(X) \mbox{sgn}(x_k - x_l) \frac{\partial}{\partial x_k}
\left[
\psi_F^* (x,X)  \psi_F (y,X)
\right]
\nonumber
\end{align}
This is obtained by collecting the corresponding terms from the first and the 
second part of the $1/c$ correction to the RSPDM in Eq. \eqref{eq:rspdmExA}.
Let us now look at the \(k\)th integration only. Integrating by parts we
have:
\begin{multline*}
\left[ \mbox{sgn}(x_k - x_l) \psi_F^* (x,X)  \psi_F (y,X) 
\right]\Big|_{-\infty}^{\infty} - \\
\int_{-\infty}^{\infty} \!\!\! dx_k \,
2\delta (x_k - x_l) \psi_F^* (x,X)  \psi_F (y,X)
\end{multline*}
The first term is zero due to boundary condition at infinity and the second
term is zero because of the antisymmetry of \(\psi_F\).
By using the method presented above it is straightforward to see that 
\[
\int_{-\infty}^{\infty} \!\! dx \,[\eta(x,x,t)+\eta(x,x,t)^*] = 0
\]
where \(\eta\) is defined in Eq. \eqref{eq:auxterm}. A
direct consequence is that the normalization constant 
${\mathcal N}(c) = 1$ in the \(1/c\) approximation cosidered here.

\section{\label{app:determinant} Determinant expansion}

In this appendix we derive Eq. \eqref{eq:eta-num}. The derivation 
is similar to that for the RSPDM of a TG gas \cite{Pezer2007}.
To simplify the notation, we will suppress the time variable 
in all formulas. 
By inserting Eq. \eqref{eq:psiFdet} for $\psi_F$ in 
Eq. \eqref{eq:auxterm} for $\eta$ we obtain
\begin{eqnarray} 
\eta(x,y) = \frac{N(N-1)}{N!}
 \sum_{ij}(-)^{i+j} [\phi_{i}^{'}(x)]^{*} \phi_{j}(y) \nonumber\\
  \sum_{P,Q} (-)^{P}(-)^{Q} \overline{I}_{p_2,q_2} I_{p_3,q_3} 
       \ldots I_{p_N,q_N}.  
\label{eq:eta1}
\end{eqnarray}
Here, $P\in S_{N-1}$ is a permutation of the $N-1$ indices 
not including $i$: $(p_2,\ldots,p_N)=P(1\ldots i-1\ i+1\ldots N)$. 
Similarly, $Q\in S_{N-1}$ permutes the $N-1$ indices 
not including $j$: $(q_2,\ldots,q_N)=Q(1\ldots j-1\ j+1\ldots N)$.

We consider the sum over the permutations. Let 
$(q_3'',\ldots,q_N'')$ be the ordered series of integers 
$\{1,\ldots,N\}\backslash\{l,j\}$, where $l\not=j$, and  
$Q'\in S_{N-2}$ a permutation of these $N-2$ numbers, 
$(q_3',\ldots,q_N')=Q'(q_3'',\ldots,q_N'')$.
With these, the sum over permutations in Eq.~\eqref{eq:eta1} can be written as
\begin{alignat}{2} 
& \frac{1}{(N-2)!}\sum_{P,Q}(-)^{P+Q} \overline{I}_{p_2,q_2} I_{p_3,q_3} 
\cdots I_{p_N,q_N} \nonumber \\ 
& = \frac{1}{(N-2)!}\sum_{\substack{l=1\\ (l\neq j)}}^{N} 
    (-)^{l+1}
    \sum_{P,Q'} (-)^{P+Q'}\,
    \overline{I}_{p_{2},l}   
    I_{p_3,q_3'} \cdots I_{p_N,q_N'} \nonumber \\
& = \sum_{\substack{l=1\\ (l\neq j)}}^{N} 
    \sum_{P} (-)^{l+1} (-)^{P}\, 
    \overline{I}_{p_{2},l}   
    I_{p_3,q_3''} \cdots I_{p_N,q_N''} \nonumber \\
& = \sum_{\substack{l=1\\ (l\neq j)}}^{N} 
    \det {\mathbf P}_{i,j}^{(l)} (x,y).
\label{eq:Qcrt}
\end{alignat}
Here, \({\mathbf P}_{i,j}^{(l)}\) is a $N-1 \times N-1$ matrix obtained by 
eliminating the $i$th row and the $j$th column from the matrix 
\({\mathbf P}^{(l)}\) defined in Eq.\eqref{eq:AmatDP}.
This yields the following result for $\eta(x,y)$:
\begin{align} 
\eta(x,y) 
 =& \sum_{ij}[\phi_{i} ^{'}(x)]^{*} \phi_{j}(y)  
    \sum_{\substack{l=1\\ (l\neq j)}} (-)^{i+j} 
       \det {\mathbf P}_{i,j}^{(l)} (x,y)  
\label{eq:eta}
\end{align}
The final result, Eq.~\eqref{eq:eta-num}, is obtained from \eqref{eq:eta} with
the use of standard matrix algebra connecting minors, the cofactor matrix, and 
the matrix inverse.

\end{document}